\documentclass[aps,prb,floatfix,twocolumn,superscriptaddress]{revtex4}
\usepackage[utf8]{inputenc}
\usepackage{mathrsfs}
\usepackage{graphicx}
\usepackage{epstopdf}
\usepackage[english]{babel}
\usepackage{amsmath}
\usepackage{amssymb}
\usepackage{xcolor}
\usepackage{relsize}
\usepackage{CJK}
\usepackage{textcomp}
\usepackage{dsfont}
\usepackage{bm}
\usepackage{float}

\newcommand{\mi}{\mathrm{i}}

\allowdisplaybreaks

\begin{document}

\title{Wilczek–Zee Realization of Uhlmann Parallel Transport}

\author{Yu-Huan Huang}
\affiliation{School of Physics, Southeast University, Jiulonghu Campus, Nanjing 211189, China}
\author{Xu-Yang Hou}
\affiliation{School of Physics, Southeast University, Jiulonghu Campus, Nanjing 211189, China}
\author{Jia-Chen Tang}
\affiliation{School of Physics, Southeast University, Jiulonghu Campus, Nanjing 211189, China}
\author{Hao Guo}
\email{guohao.ph@seu.edu.cn}
\affiliation{School of Physics, Southeast University, Jiulonghu Campus, Nanjing 211189, China}
\affiliation{Hefei National Laboratory, University of Science and Technology of China, Hefei 230088, China}

\begin{abstract}
The Uhlmann phase extends geometric phases to mixed quantum states via a parallel-transport condition on purification amplitudes, yet its direct implementation under standard Hamiltonian dynamics is obstructed by the non-Hermitian nature of the purification. We establish that for any smooth one-dimensional closed loop of full-rank qubit density matrices, there exists a four-level Hermitian parent Hamiltonian whose doubly degenerate ground-state subspace carries a Wilczek--Zee connection exactly equal to the Uhlmann connection. Consequently, the Uhlmann holonomy is faithfully reproduced by adiabatic evolution in the enlarged system. We further prove that this auxiliary-field construction is obstructed in generic two-dimensional parameter spaces by a Frobenius integrability condition, which we derive explicitly. The one-dimensional Uhlmann phase is thus placed on the same footing as the non-Abelian Berry phase, offering a purely Hermitian, Hamiltonian-based route to simulating mixed-state geometric phases. Numerical integration of the adiabatic dynamics confirms the exact correspondence and validates the convergence to the Uhlmann holonomy in the large-gap limit.
\end{abstract}

\maketitle

\section{Introduction}

Geometric phases are fundamental to the topological characterization of quantum systems~\cite{Bohm_GPbook,ChruscinskiBook}. The Berry phase, acquired during adiabatic cyclic evolution of pure states, underpins the modern classification of topological insulators and superconductors through the Berry curvature and related invariants~\cite{Berry1984,TKNN,Haldane,KaneRMP,ZhangSCRMP,ChiuRMP}. Its non-Abelian generalization, the Wilczek--Zee (WZ) phase~\cite{PhysRevLett.52.2111}, arises when a degenerate subspace undergoes adiabatic evolution, and has found applications in holonomic quantum computation. These pure-state geometric phases have been observed in a wide range of platforms, including superconducting qubits, ultracold atoms, and photonic systems~\cite{Leek2007,Atala2013,Wang2023}.

For mixed states, the mathematically rigorous extension is provided by the Uhlmann phase~\cite{Uhlmann86,Uhlmann91}, which is constructed from the parallel-transport condition imposed on the purification amplitude $W$ of a density matrix $\rho=WW^\dagger$. This framework has been widely applied to the study of topological phenomena at finite temperature and in open quantum systems~\cite{Huang_2014,Viyuela2015,PhysRevLett.112.130401,PhysRevLett.119.015702}. An interferometric formulation of mixed-state geometric phases was developed by Sj\"{o}qvist \textit{et al.} and demonstrated in NMR experiments~\cite{PhysRevLett.85.2845,PhysRevLett.91.100403}. The Uhlmann phase itself has been observed in superconducting circuits, photonic quantum walks, and programmable quantum processors~\cite{Viyuela2018,mastandrea2025,wang2025}; in those experiments, however, the phase was extracted from interferometric or tomographic measurements on the density matrix, without implementing the parallel transport of the purification at the dynamical level.

Despite its conceptual importance, it has been shown that the Uhlmann parallel-transport condition is fundamentally incompatible with standard Hermitian Hamiltonian dynamics~\cite{P1}. The root of this obstruction lies in the fact that the purification amplitude $W$ is generally a non-Hermitian matrix, so its evolution under a physical Hamiltonian $H$ does not preserve the Uhlmann condition $W^\dagger\dot W=\dot W^\dagger W$. In one approach to circumvent this difficulty, classical electrical-circuit emulators have been developed to map the vectorized Uhlmann dynamics onto Kirchhoff equations~\cite{Huang2026eleCSU}, offering an accessible classical analogue that reproduces the Uhlmann geometric phase and its topological transition. In the present work, we pursue a complementary strategy that operates entirely within quantum Hamiltonian dynamics: inspired by the concept of a parent Hamiltonian introduced by Yang \textit{et al.}~\cite{PhysRevLett.130.220401} for non-Hermitian systems, we construct a Hermitian \textit{parent Hamiltonian} acting on an enlarged Hilbert space, such that the Uhlmann parallel transport of the original mixed state is mapped to the standard unitary evolution of a pure state in the larger system. This idea lifts the problem from the purification space to the familiar terrain of Hermitian adiabatic dynamics.

In this work, we prove that such a realization is indeed possible for any smooth one-dimensional full-rank qubit loop. Our central idea is to lift the Uhlmann parallel-transport condition into the Wilczek--Zee connection of a doubly degenerate subspace in a higher-dimensional pure-state system. By adiabatically evolving within this degenerate subspace, one naturally generates the Uhlmann holonomy, thereby translating the mixed-state geometric phase into standard unitary Hamiltonian dynamics at the price of doubling the Hilbert-space dimension. This correspondence provides an affirmative answer to the question of whether Uhlmann parallel transport can be simulated by standard Hamiltonian dynamics: it can for one-dimensional loops, provided one is willing to enlarge the Hilbert space and engineer the parent Hamiltonian accordingly. Numerical integration of the time-dependent Schr\"odinger equation confirms that the Uhlmann holonomy is faithfully reproduced once the adiabatic condition is satisfied.

We further demonstrate that this correspondence is fundamentally restricted to one-dimensional parameter spaces by the present auxiliary-field construction. For a two-dimensional parameter manifold, the auxiliary matrix $V(t)$ that underlies our construction must satisfy a Frobenius integrability condition that is generically violated by the Uhlmann curvature. We derive the explicit obstruction formula and explain why the one-dimensional case escapes this obstruction trivially. Alongside the analytical results, numerical simulations of the adiabatic dynamics are presented to corroborate the exact correspondence in the tractable one-dimensional case.

The remainder of the paper is organized as follows. Section~\ref{Sec2} reviews the Uhlmann parallel-transport condition and the Uhlmann connection for qubit density matrices, then presents the main theorem and its proof. Section~\ref{example} provides the explicit analytic construction for the equatorial qubit loop and verifies the WZ–Uhlmann matching. Section~\ref{Secobstruction} derives the curvature obstruction that prevents extending the auxiliary-field construction to higher-dimensional parameter spaces. Section~\ref{discussion} discusses experimental feasibility, generalizations to higher-rank density matrices, and comparisons with non-Hermitian schemes. Section~\ref{conclusion} concludes with a summary and outlook.

\section{Theoretical Framework}\label{Sec2}
\subsection{Uhlmann Parallel Transport and Connection}
\label{Uhlmann}

We focus on a two-level system, as the construction presented in this work applies to qubit density matrices. Consider a smooth closed loop of full-rank density matrices traversing a one-dimensional parameter space,
\begin{equation}
\rho(t),\quad t\in[0,\tau],\quad \rho(0)=\rho(\tau),
\end{equation}
with instantaneous spectral decomposition
\begin{equation}
\rho(t)=\lambda_+(t)|+(t)\rangle\langle +(t)|+\lambda_-(t)|-(t)\rangle\langle -(t)|,
\end{equation}
where $0<\lambda_-(t)\le\lambda_+(t)<1$ and $\lambda_++\lambda_-=1$. The square-root amplitudes are
\begin{equation}
a(t)=\sqrt{\lambda_+(t)},\quad b(t)=\sqrt{\lambda_-(t)},\quad a^2+b^2=1.
\end{equation}

The purification of $\rho$ is an invertible $2\times 2$ matrix $W$ satisfying $\rho=WW^\dagger$. Since $\rho$ is full-rank, $W$ admits the unique polar decomposition $W=\sqrt{\rho}\,U$ with $U\in U(2)$, where $U$ embodies the $U(2)$ gauge freedom of the purification. The Uhlmann parallel-transport condition,
\begin{equation}
W^\dagger\dot W=\dot W^\dagger W,
\end{equation}
minimizes the Hilbert-Schmidt distance between infinitesimally separated purifications, making the distance independent of the gauge choice $U$~\cite{Uhlmann86}. By substituting $W=\sqrt{\rho}\,U$ into this equation, one finds that the unitary factor $U$ undergoes a parallel transport $\dot U+\mathcal A_{\text{U}} U=0$ governed by the Uhlmann connection $\mathcal A_{\text{U}}$:
\begin{equation}
\mathcal{A}_{\text{U}} = - \sum_{ij=\pm}
\frac{\langle i| [\mathrm{d}\sqrt{\rho}, \sqrt{\rho}] |j\rangle}
{\lambda_i + \lambda_j}
|i\rangle\langle j| .
\label{AU}
\end{equation}
It follows directly from Eq.~(\ref{AU}) that $\mathcal A_{\text{U}}^\dagger=-\mathcal A_{\text{U}}$, a property that will be essential for the construction below.

For a cyclic evolution, the Uhlmann holonomy is
\begin{equation}
U_{\text{U}}(\tau)=\mathcal P\exp\!\left(-\int_0^\tau\mathcal A_{\text{U}}(t)\,\mathrm dt\right),
\end{equation}
with $\mathcal{P}$ the path-ordering operator. The Uhlmann geometric phase is given by
\begin{equation}
\Phi_{\text{U}}=\arg\operatorname{Tr}\!\bigl[\rho(0)\,U_{\text{U}}(\tau)\bigr].
\end{equation}
Since $\rho$ is full-rank, $b(t)>0$ for all $t\in[0,\tau]$, which ensures that the matrix $\mathcal A_{\text{U}}(t)/b^2(t)$ is smooth and finite, a fact that plays a crucial role in what follows.

\subsection{Exact Uhlmann--Wilczek--Zee Correspondence for One-Dimensional Loops}
\label{theorem}

With the Uhlmann connection at hand, we now state and prove the main result. The restriction to one-dimensional parameter spaces is not merely a simplification; as we shall see later in Eq.~\ref{obstruction}, it is a necessary condition for the present construction to hold.

\textbf{Theorem 1.}
Let $\rho(t)$, $t\in[0,\tau]$, be a smooth closed loop of full-rank qubit density matrices traversing a one-dimensional parameter space, with Uhlmann connection $\mathcal A_{\text{U}}(t)$. Then there exists a smooth four-level Hermitian Hamiltonian $H(t)$ possessing a doubly-degenerate eigenspace such that the corresponding Wilczek--Zee connection satisfies
\begin{equation}
\mathcal A_{\rm WZ}(t)=\mathcal A_{\text{U}}(t)
\end{equation}
along the entire loop. Consequently, $U_{\rm WZ}(\tau)=U_{\text{U}}(\tau)$, and the Uhlmann holonomy is reproduced exactly by adiabatic evolution in the parent system.

\textbf{Proof:}
The proof proceeds in four steps.

\paragraph*{Step 1: Auxiliary unitary matrix $V(t)$.}
To embed the Uhlmann connection into a larger pure-state system, we introduce an auxiliary unitary matrix $V(t)$ as the unique smooth solution of the linear matrix differential equation
\begin{equation}
\label{Vode}
\frac{\mathrm dV}{\mathrm dt}=V(t)\,\frac{\mathcal A_{\text{U}}(t)}{b^2(t)},\quad V(0)=I_2,
\end{equation}
which is well defined because the parameter space is one-dimensional and $b(t)>0$ guarantees smoothness of the coefficient matrix. To verify unitarity, we compute
\begin{align}
\frac{\mathrm d}{\mathrm dt}\bigl(VV^\dagger\bigr)
&=V\frac{\mathcal A_{\text{U}}}{b^2}V^\dagger+V\left(\frac{\mathcal A_{\text{U}}}{b^2}\right)^\dagger V^\dagger =0,
\end{align}
using $\mathcal A_{\text{U}}^\dagger=-\mathcal A_{\text{U}}$ and the reality of $b^2$. Hence $VV^\dagger$ is constant; the initial condition $V(0)=I_2$ then gives $V(t)V^\dagger(t)=I_2$ for all $t$. For a finite-dimensional square matrix this implies $V^\dagger V=I$, so $V(t)\in U(2)$ for all $t\in[0,\tau]$.

\paragraph*{Step 2: Orthonormal frame in the auxiliary subspace.}
The construction is basis-independent, but for concreteness we fix an orthonormal basis $\{|1\rangle,|2\rangle,|3\rangle,|4\rangle\}$ of $\mathbb C^4$ and designate the two-dimensional auxiliary subspace as $\mathcal H_{\rm aux}=\mathrm{span}\{|2\rangle,|3\rangle\}$. We regard $V(t)$ as a unitary operator on $\mathcal H_{\rm aux}$ with matrix elements $V_{mn}(t)$ in the basis $\{|2\rangle,|3\rangle\}$. Define
\begin{equation}
|\chi_m(t)\rangle=\sum_{n=1}^{2}V_{nm}(t)\,|n+1\rangle,\quad m=1,2.
\label{chi_def}
\end{equation}
Because $V(t)$ is unitary on $\mathcal H_{\rm aux}$, these states are orthonormal:
\begin{equation}
\langle\chi_m|\chi_n\rangle=\sum_{\alpha=1}^{2}V_{\alpha m}^*V_{\alpha n}=(V^\dagger V)_{mn}=\delta_{mn}.
\end{equation}
Their overlap derivatives are
\begin{align}
\langle\chi_m|\dot\chi_n\rangle=\sum_{\alpha=1}^{2}V_{\alpha m}^*\dot V_{\alpha n}
=(V^\dagger\dot V)_{mn}=\frac{(\mathcal A_{\text{U}})_{mn}}{b^2},
\label{chi_overlap}
\end{align}
where the last equality follows from Eq.~(\ref{Vode}).

\paragraph*{Step 3: Constructing the degenerate pure states.}
Now define
\begin{align}
|\psi_1(t)\rangle&=a(t)\,|1\rangle+b(t)\,|\chi_1(t)\rangle,\notag\\
|\psi_2(t)\rangle&=a(t)\,|4\rangle+b(t)\,|\chi_2(t)\rangle.
\end{align}
Using $a^2+b^2=1$ and the orthonormality of $\{|\chi_m\rangle\}$, we obtain
\begin{align}
\langle\psi_1|\psi_1\rangle&=a^2+b^2\langle\chi_1|\chi_1\rangle=1,\notag\\
\langle\psi_2|\psi_2\rangle&=a^2+b^2\langle\chi_2|\chi_2\rangle=1,\notag\\
\langle\psi_1|\psi_2\rangle&=a^2\langle 1|4\rangle+b^2\langle\chi_1|\chi_2\rangle=0.
\end{align}
Thus $\{|\psi_1\rangle,|\psi_2\rangle\}$ forms an orthonormal basis of a smooth two-dimensional subbundle of $\mathbb C^4$.
The rank-two projector onto this subspace is
\begin{equation}
P(t)=|\psi_1(t)\rangle\langle\psi_1(t)|+|\psi_2(t)\rangle\langle\psi_2(t)|.
\end{equation}
We define the parent Hamiltonian
\begin{equation}
H(t)=\Delta\bigl[I-P(t)\bigr],\quad \Delta>0.
\end{equation}
By construction $H(t)$ is Hermitian. Since $P(t)$ is a projector with eigenvalues $1$ (doubly degenerate, on the span of $|\psi_1\rangle,|\psi_2\rangle$) and $0$ (doubly degenerate, on the orthogonal complement), the spectrum of $H(t)$ consists of a doubly-degenerate ground-state eigenvalue $0$ and a doubly-degenerate excited-state eigenvalue $\Delta>0$. The finite gap $\Delta$ guarantees that the adiabatic theorem applies, as will be discussed in Sec.~\ref{parent}.

\paragraph*{Step 4: Matching the Wilczek--Zee and Uhlmann connections.}
The Wilczek--Zee connection is defined by
\begin{equation}
(\mathcal A_{\rm WZ})_{mn}=\langle\psi_m|\dot\psi_n\rangle.
\end{equation}
For the diagonal element $m=n=1$,
\begin{align}
\langle\psi_1|\dot\psi_1\rangle&=a\dot a+\dot a b\langle 1|\chi_1\rangle+a\dot b\langle\chi_1|1\rangle\notag\\&\quad+b\dot b\langle\chi_1|\chi_1\rangle+b^2\langle\chi_1|\dot\chi_1\rangle\nonumber\\
&=a\dot a+b\dot b+b^2\langle\chi_1|\dot\chi_1\rangle.
\end{align}
Since $a^2+b^2=1$, we have $a\dot a+b\dot b=0$, yielding
\begin{equation}
\langle\psi_1|\dot\psi_1\rangle=b^2\langle\chi_1|\dot\chi_1\rangle=(\mathcal A_{\text{U}})_{11},
\end{equation}
where Eq.~(\ref{chi_overlap}) was used. Similarly,
\begin{equation}
\langle\psi_2|\dot\psi_2\rangle=(\mathcal A_{\text{U}})_{22}.
\end{equation}
For the off-diagonal element,
\begin{align}
\langle\psi_1|\dot\psi_2\rangle&=a\dot a\langle 1|4\rangle+\dot a b\langle 1|\chi_2\rangle+a\dot b\langle\chi_1|4\rangle\nonumber\\
&\quad+b\dot b\langle\chi_1|\chi_2\rangle+b^2\langle\chi_1|\dot\chi_2\rangle.
\end{align}
Because $|1\rangle$ and $|4\rangle$ are orthogonal to $\mathcal H_{\rm aux}$ and $\langle\chi_1|\chi_2\rangle=0$, all terms except the last vanish:
\begin{equation}
\langle\psi_1|\dot\psi_2\rangle=b^2\langle\chi_1|\dot\chi_2\rangle=(\mathcal A_{\text{U}})_{12}.
\end{equation}
Likewise,
\begin{equation}
\langle\psi_2|\dot\psi_1\rangle=(\mathcal A_{\text{U}})_{21}.
\end{equation}
Collecting these results, we obtain
\begin{equation}
\mathcal A_{\rm WZ}(t)=\mathcal A_{\text{U}}(t).
\end{equation}
Since the connections are identical, their path-ordered exponentials coincide,
\begin{equation}
U_{\rm WZ}(\tau)=\mathcal P\exp\!\left(-\int_0^\tau\mathcal A_{\rm WZ}\,\mathrm dt\right)=U_{\text{U}}(\tau),
\end{equation}
which establishes the theorem.

Whether four dimensions are strictly necessary or a lower-dimensional embedding exists is a natural question; a brief discussion is given in Appendix~\ref{app1}.
A different connection between the Uhlmann and Wilczek--Zee phases was explored in our previous work~\cite{bdpw-856t}, where the two phases were compared in a fixed four-level model. In contrast, Theorem~1 provides a constructive equivalence: we engineer a parent Hamiltonian whose Wilczek--Zee connection exactly reproduces the Uhlmann connection of an arbitrary qubit loop.

\subsection{Parent Hamiltonian and Adiabatic Simulation}
\label{parent}

Theorem~1 guarantees the existence of a Hermitian parent Hamiltonian whose degenerate ground-state manifold carries the Uhlmann connection. We now discuss how this enables Hamiltonian simulation of Uhlmann parallel transport.

\subsubsection{Adiabatic evolution in the degenerate subspace}

Theorem~1 establishes the kinematic equivalence between the Uhlmann and Wilczek--Zee connections. To turn this into a dynamical simulation, we now consider adiabatic evolution under the parent Hamiltonian. Suppose the four-level system is prepared in a state $|\Psi(0)\rangle$ lying entirely within the ground-state subspace of $H(0)$:
\begin{equation}
|\Psi(0)\rangle=c_1|\psi_1(0)\rangle+c_2|\psi_2(0)\rangle,\quad |c_1|^2+|c_2|^2=1.
\end{equation}
Under the time-dependent Schr\"odinger equation
\begin{equation}
i\frac{\mathrm d}{\mathrm dt}|\Psi(t)\rangle=H(t)|\Psi(t)\rangle,
\label{Schr}
\end{equation}
with $H(t)=\Delta(I-P(t))$, the adiabatic theorem for degenerate states~\cite{PhysRevLett.52.2111,Kato1950} guarantees that if the gap $\Delta$ is large compared to the rate of change of the projector,
\begin{equation}
\Delta\gg \max_{t\in[0,\tau]}\|\dot P(t)\|,
\end{equation}
then the state remains confined to the instantaneous ground-state subspace. Within this subspace, the coefficients evolve according to the Wilczek--Zee parallel-transport equation
\begin{equation}
\frac{\mathrm dc_m}{\mathrm dt}=-\sum_{n=1}^{2}(\mathcal A_{\rm WZ})_{mn}\,c_n.
\end{equation}
Since $\mathcal A_{\rm WZ}=\mathcal A_{\text{U}}$ by Theorem~1, the state acquires exactly the Uhlmann holonomy after one cycle:
\begin{equation}
|\Psi(\tau)\rangle=\sum_{m,n}(U_{\text{U}})_{mn}\,c_n|\psi_m(\tau)\rangle.
\label{state_evolution}
\end{equation}
Thus, provided the evolution is adiabatic and the initial state lies in the ground-state manifold, the Uhlmann parallel transport of the original mixed-state qubit loop is faithfully reproduced by the standard Hamiltonian dynamics of the four-level parent system.

\subsubsection{Extraction of the Uhlmann phase}

To extract the Uhlmann holonomy from the parent system, one prepares the initial state in the degenerate subspace, $|\Psi(0)\rangle=|\psi_1(0)\rangle$, and lets it evolve adiabatically. After one cycle,
\begin{equation}
|\Psi(\tau)\rangle=(U_{\text{U}})_{11}|\psi_1(\tau)\rangle+(U_{\text{U}})_{21}|\psi_2(\tau)\rangle.
\end{equation}
In general, the basis $\{|\psi_m(t)\rangle\}$ constructed in Theorem~1 need not be periodic, because the auxiliary matrix $V(t)$ satisfies $\dot V=V\,\mathcal A_{\text{U}}/b^2$ with $V(0)=I$, so $|\psi_m(\tau)\rangle=|\psi_m(0)\rangle$ holds only when
\begin{equation}
\mathcal P\exp\!\int_0^\tau \frac{\mathcal A_{\text{U}}}{b^2}\,\mathrm dt = I.
\label{Vperiodic}
\end{equation}
When this condition is not satisfied, the overlap $\langle\Psi(0)|\Psi(\tau)\rangle$ involves both $(U_{\text{U}})_{11}$ and the basis overlaps $\langle\psi_m(0)|\psi_n(\tau)\rangle$, and extracting the full holonomy matrix requires quantum-state tomography or multiple interferometric measurements with different initial states.

The situation simplifies considerably when the parameter path and evolution duration are chosen such that Eq.~(\ref{Vperiodic}) holds. In this case $|\psi_m(\tau)\rangle=|\psi_m(0)\rangle$, and the interferometric overlap reduces to
\begin{equation}
\langle\Psi(0)|\Psi(\tau)\rangle=(U_{\text{U}})_{11},
\end{equation}
yielding a specific matrix element of the holonomy from a single measurement. For the equatorial loop discussed in Sec.~\ref{example}, the condition Eq.~(\ref{Vperiodic}) is satisfied when $\gamma$ is an integer or, more practically, when the system evolves over $N$ cycles such that $N\gamma\in\mathbb Z$, which can be arranged by tuning the purity $r$ or the number of cycles. In such simplified scenarios, the Uhlmann phase can be read out directly from a single interferometric signal without tomographic reconstruction.

The Uhlmann phase is defined as $\Phi_{\text{U}}=\arg\mathrm{Tr}[\rho(0)U_{\text{U}}]$. In general, a single matrix element $(U_{\text{U}})_{11}$ is not sufficient to determine this trace, because $\mathrm{Tr}[\rho(0)U_{\text{U}}]=\lambda_+(U_{\text{U}})_{++}+\lambda_-(U_{\text{U}})_{--}$ involves the diagonal elements of $U_{\text{U}}$ in the eigenbasis of $\rho(0)$, and the parent-system basis need not coincide with that eigenbasis. However, in the simplified scenarios where $V(\tau)=I$ and the parent-system basis aligns with the eigenbasis of $\rho(0)$, the two bases coincide, making the single measurement sufficient to determine $\Phi_{\text{U}}$.

\section{Explicit Construction for the Equatorial Loop}
\label{example}

Having established the general correspondence in Theorem~1, we now illustrate the construction explicitly for the equatorial qubit loop, the paradigmatic example that exhibits the topological transition of the Uhlmann phase. We follow the procedure of Theorem~1 step by step. The density matrix is
\begin{equation}
\rho(\phi)=\frac12\bigl(I+r\cos\phi\,\sigma_x+r\sin\phi\,\sigma_y\bigr),\quad \phi\in[0,2\pi],
\end{equation}
which describes a mixed state whose Bloch vector sweeps the equator of the Bloch sphere. The eigenvalues are $\lambda_\pm=\frac12(1\pm r)$, so
\begin{equation}
a=\sqrt{\frac{1+r}{2}},\quad b=\sqrt{\frac{1-r}{2}}.
\end{equation}
The parameter $\beta$ controls the deviation from the pure-state limit and will govern the Uhlmann phase below. Using Eq.~(\ref{AU}), a straightforward evaluation gives the Uhlmann connection
\begin{equation}
\mathcal A_{\text{U}}=-\frac{\mi\beta}{2}\sigma_y\,\mathrm d\phi,
\end{equation}
where $\beta=1-\sqrt{1-r^2}$.
The connection is purely off-diagonal in the eigenbasis of $\rho(0)$, reflecting the equatorial geometry.

\subsection{Solution for the auxiliary matrix $V(\phi)$}

Equation~(\ref{Vode}) becomes
\begin{equation}
\frac{\mathrm dV}{\mathrm d\phi}=V\,\frac{-\frac{\mi\beta}{2}\sigma_y}{b^2}=V\,\left(-\mi\gamma\sigma_y\right),
\end{equation}
where we have introduced the constant
\begin{equation}
\gamma\equiv\frac{\beta}{2b^2}=\frac{1-\sqrt{1-r^2}}{1-r}=\frac{r^2}{(1-r)\bigl(1+\sqrt{1-r^2}\bigr)}.
\end{equation}
The parameter $\gamma$ encodes how the purity $r$ dictates the rotation rate of the auxiliary frame. Since the coefficient matrix is constant, the solution is
\begin{equation}
V(\phi)=\exp\!\left(-\mi\gamma\phi\,\sigma_y\right)
=\begin{pmatrix}
\cos(\gamma\phi) & -\sin(\gamma\phi)\\[4pt]
\sin(\gamma\phi) & \cos(\gamma\phi)
\end{pmatrix}.
\end{equation}
Geometrically, $V(\phi)$ describes a uniform rotation of the auxiliary frame as the parameter $\phi$ advances. Unitarity is manifest because the exponent is anti-Hermitian.

\subsection{The orthonormal frame and degenerate pure states}

According to Eq.~(\ref{chi_def}), we construct the auxiliary frame vectors
\begin{align}
|\chi_1(\phi)\rangle
&=\cos(\gamma\phi)\,|2\rangle+\sin(\gamma\phi)\,|3\rangle,\notag\\
|\chi_2(\phi)\rangle
&=-\sin(\gamma\phi)\,|2\rangle+\cos(\gamma\phi)\,|3\rangle.
\end{align}
These two states form an orthonormal basis of the auxiliary subspace that rotates with angular velocity $\gamma$ as $\phi$ varies. One readily verifies $\langle\chi_m|\chi_n\rangle=\delta_{mn}$ and, using Eq.~(\ref{chi_overlap}),
\begin{align}
\langle\chi_1|\partial_\phi|\chi_2\rangle&=(V^\dagger\dot V)_{12}=-\gamma,\notag\\
\langle\chi_2|\partial_\phi|\chi_1\rangle&=(V^\dagger\dot V)_{21}=+\gamma.
\end{align}
Now, following Step 3 of the theorem, we build the degenerate pure states by mixing the fixed basis states with the auxiliary frame:
\begin{align}
|\psi_1(\phi)\rangle&=a\,|1\rangle+b\,|\chi_1(\phi)\rangle\nonumber\\
&=a\,|1\rangle+b\cos(\gamma\phi)\,|2\rangle+b\sin(\gamma\phi)\,|3\rangle,\notag\\
|\psi_2(\phi)\rangle&=a\,|4\rangle+b\,|\chi_2(\phi)\rangle\nonumber\\
&=a\,|4\rangle-b\sin(\gamma\phi)\,|2\rangle+b\cos(\gamma\phi)\,|3\rangle.
\label{initial_state}
\end{align}
The amplitudes $a$ and $b$ control the relative weight between the fixed and rotating components, with the pure-state limit $r\to 1$ corresponding to $a\to 1$, $b\to 0$.

\subsection{Verification of the Wilczek--Zee--Uhlmann matching}

We compute the Wilczek--Zee connection directly from the states above. The diagonal elements vanish:
\begin{align}
\langle\psi_1|\partial_\phi|\psi_1\rangle&=a\partial_\phi a+b^2\langle\chi_1|\partial_\phi|\chi_1\rangle=0,\notag\\
\langle\psi_2|\partial_\phi|\psi_2\rangle&=a\partial_\phi a+b^2\langle\chi_2|\partial_\phi|\chi_2\rangle=0,
\end{align}
since $a^2+b^2=1$ implies $a\partial_\phi a+b\partial_\phi b=0$ and the auxiliary frame has vanishing diagonal overlaps. The off-diagonal elements are
\begin{align}
\langle\psi_1|\partial_\phi|\psi_2\rangle&=b^2\langle\chi_1|\partial_\phi|\chi_2\rangle=-b^2\gamma=-\frac{\beta}{2},\notag\\
\langle\psi_2|\partial_\phi|\psi_1\rangle&=b^2\langle\chi_2|\partial_\phi|\chi_1\rangle=+b^2\gamma=+\frac{\beta}{2}.
\end{align}
Therefore,
\begin{equation}
\mathcal A_{\rm WZ}=
\begin{pmatrix}
0 & -\frac{\beta}{2}\\[4pt]
\frac{\beta}{2} & 0
\end{pmatrix}
\mathrm d\phi
=-\frac{\mi\beta}{2}\sigma_y\,\mathrm d\phi
=\mathcal A_{\text{U}},
\end{equation}
confirming that the explicit construction reproduces the Uhlmann connection exactly, as guaranteed by Theorem~1.

The Uhlmann holonomy for a full cycle follows immediately:
\begin{align}
U_{\text{U}}(2\pi)&=\mathcal P\exp\!\left(-\oint\mathcal A_{\text{U}}\right)=\exp(\mi\pi\beta\sigma_y)\notag\\&=\cos(\pi\beta)I+\mi\sin(\pi\beta)\sigma_y.
\end{align}
Since $\rho(0)=\frac12(I+r\sigma_x)$, we obtain the Uhlmann phase
\begin{equation}
\Phi_{\text{U}}=\arg\mathrm{Tr}[\rho(0)U_{\text{U}}(2\pi)]=\arg\left[\cos(\pi\beta)\right],
\end{equation}
which exhibits the well-known topological transition: $\Phi_{\text{U}}=0$ for $\beta<1/2$ and $\Phi_{\text{U}}=\pi$ for $\beta>1/2$, corresponding to the critical purity $r_c=\sqrt{3}/2$.

\subsection{Parent Hamiltonian in matrix form}

The projector $P(\phi)=|\psi_1\rangle\langle\psi_1|+|\psi_2\rangle\langle\psi_2|$ can be written explicitly in the $\{|1\rangle,|2\rangle,|3\rangle,|4\rangle\}$ basis. After a straightforward calculation,
\begin{align}
&P(\phi)\notag\\=&
\begin{pmatrix}
a^2 & ab\cos(\gamma\phi) & ab\sin(\gamma\phi) & 0\\[4pt]
ab\cos(\gamma\phi) & b^2 & 0 & -ab\sin(\gamma\phi)\\[4pt]
ab\sin(\gamma\phi) & 0 & b^2 & ab\cos(\gamma\phi)\\[4pt]
0 & -ab\sin(\gamma\phi) & ab\cos(\gamma\phi) & a^2
\end{pmatrix}.
\label{P(t)}
\end{align}
The parent Hamiltonian is $H(\phi)=\Delta(I-P(\phi))$. Its matrix elements are
\begin{align}
H_{11}&=H_{44}=\Delta b^2=\Delta\frac{1-r}{2},\notag\\
H_{22}&=H_{33}=\Delta a^2=\Delta\frac{1+r}{2},\notag\\
H_{12}&=H_{21}^*=-\Delta ab\cos(\gamma\phi),\notag\\
H_{13}&=H_{31}^*=-\Delta ab\sin(\gamma\phi),\notag\\
H_{24}&=H_{42}^*=\Delta ab\sin(\gamma\phi),\notag\\
H_{34}&=H_{43}^*=-\Delta ab\cos(\gamma\phi),
\label{parent Hamiltonian}
cx\end{align}
with all other off-diagonal elements vanishing. This Hamiltonian is manifestly Hermitian and $\phi$-dependent. Physically, the diagonal terms represent level detunings, while the off-diagonal terms describe couplings that are modulated harmonically as the parameter sweeps the loop.

The gap is uniform: the two lowest eigenvalues are $0$ (degenerate), and the two excited eigenvalues are $\Delta$ (degenerate). The adiabatic condition requires
\begin{equation}
\Delta\gg \omega\max_\phi\|\partial_\phi P(\phi)\|,
\end{equation}
where $\omega=\dot\phi$ is the angular velocity of the parameter sweep. For the equatorial loop, $\max_\phi\|\partial_\phi P(\phi)\|=ab\gamma$, yielding
\begin{equation}
\Delta\gg \omega ab\gamma = \omega\frac{\sqrt{1-r^2}}{2}\frac{r^2}{(1-r)\bigl(1+\sqrt{1-r^2}\bigr)}.
\label{adiabatic}
\end{equation}
Thus, for a given purity $r$ and sweep rate $\omega$, choosing $\Delta$ sufficiently large ensures adiabaticity. The required gap diverges as $r\to 1$, reflecting the increasing fragility of the construction near the pure-state limit.

\paragraph*{Pure-state limit.}
As $r\to 1$, the coefficient $\gamma=\beta/(2b^2)$ diverges. Consequently, the auxiliary ODE coefficient $\mathcal A_{\text{U}}/b^2=-\mi\gamma\sigma_y$ in Eq.~(\ref{Vode}) diverges, and the proof of Theorem~1 no longer applies at $r=1$. It is worth noting that the physical Wilczek--Zee connection $\mathcal A_{\rm WZ}=\mathcal A_{\text{U}}$ itself remains finite; the divergence is an artifact of the particular auxiliary-field construction, not of the underlying geometry. A separate asymptotic analysis of the parent Hamiltonian near $r=1$ is required to assess whether an alternative construction might bridge the pure-state limit.

\subsection{Numerical verification of adiabatic convergence}

To verify that the parent Hamiltonian reproduces the Uhlmann holonomy through standard adiabatic evolution, we numerically integrate the time-dependent Schr\"odinger equation~(\ref{Schr}) with the Hamiltonian~(\ref{parent Hamiltonian}).

The system is prepared in one of the degenerate ground states, $|\Psi(0)\rangle=|\psi_j(0)\rangle$ ($j=1,2$), given by Eq.~(\ref{initial_state}). After one full cycle, the evolved state is projected onto the instantaneous basis $\{|\psi_1(\tau)\rangle,|\psi_2(\tau)\rangle\}$ of the degenerate subspace at $t=\tau$. According to Eq.~(\ref{state_evolution}), the matrix elements of the Uhlmann holonomy are extracted as
\begin{equation}
(U_{\rm U})_{mj}=\langle\psi_m(\tau)|\Psi_j(\tau)\rangle,\qquad m,j=1,2.
\end{equation}
The Uhlmann phase is then evaluated from the reconstructed holonomy,
\begin{equation}
\Phi_{\rm U}=\arg\operatorname{Tr}\!\bigl[\rho(0)\,U_{\rm U}\bigr].
\end{equation}
We fix the sweep period to $T=2\pi$ (so $\omega=1$) and vary the dimensionless gap $\Delta/\omega$.

\begin{figure}[ht]
    \centering
    \includegraphics[width=3.4in]{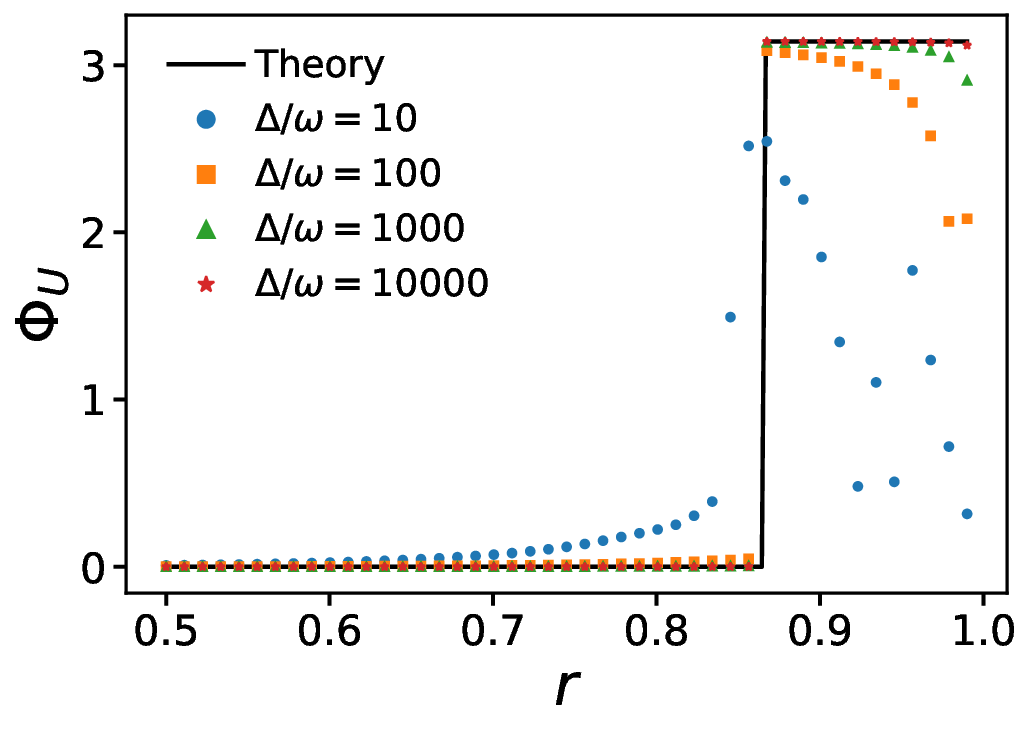}
    \caption{Uhlmann phase $\Phi_{\rm U}$ versus the purity parameter $r$. The solid black line is the theoretical prediction, while the symbols show numerical results obtained from adiabatic evolution under the parent Hamiltonian for $\Delta/\omega=10$ (circles), $100$ (squares), $1000$ (triangles), and $10000$ (stars).}
    \label{fig:phase_results}
\end{figure}

The numerical results are summarized in Fig.~\ref{fig:phase_results}. The adiabatic condition~(\ref{adiabatic}) is most stringent near $r\to 1$, where the effective driving rate $ab\gamma$ reaches its maximum. Quantitatively, $ab\gamma$ diverges as $1/\sqrt{1-r}$ in the pure-state limit, so that the required gap $\Delta/\omega$ must grow inversely with the distance to $r=1$. For instance, $ab\gamma\approx2.5$ at $r=0.96$, whereas it increases to approximately $6$ at $r=0.99$. Consequently, $\Delta/\omega=10$ is insufficient throughout the entire range, and even $\Delta/\omega=100$, though adequate near $r_c$, fails to suppress nonadiabatic transitions at larger $r$, leading to the visible deviations seen in Fig.~\ref{fig:phase_results}. As the gap is increased to $\Delta/\omega=1000$ and $10000$, the numerical data progressively collapse onto the theoretical curve, with $\Delta/\omega=10000$ essentially indistinguishable from the exact result. This systematic convergence confirms that the parent Hamiltonian faithfully reproduces the Uhlmann holonomy in the adiabatic limit, validating the correspondence established in Theorem~1, while the increasing difficulty near $r\to 1$ corroborates the pure-state-limit obstruction discussed in Sec.~\ref{example}.

\section{Curvature Obstruction in Higher-Dimensional Parameter Spaces}
\label{Secobstruction}

Theorem~1 establishes the correspondence for one-dimensional loops, but the restriction to one dimension is not a technical limitation of the proof; it reflects a genuine geometric obstruction. We now show that the auxiliary-field construction cannot be extended to generic two-dimensional parameter manifolds.

Consider a two-dimensional parameter space with coordinates $(\lambda^1,\lambda^2)$. The Uhlmann connection is a matrix-valued 1-form $\mathcal A_{\text{U}}=\mathcal A_{U,\mu}\,\mathrm d\lambda^\mu$. To generalize the construction of Sec.~\ref{theorem}, one would need a matrix-valued function $V(\lambda^1,\lambda^2)\in U(2)$ satisfying
\begin{equation}
\partial_\mu V = V B_\mu,\quad B_\mu\equiv\frac{\mathcal A_{U,\mu}}{b^2}.
\label{2d_V}
\end{equation}
The Frobenius integrability condition for this system is
\begin{equation}
\partial_\mu B_\nu-\partial_\nu B_\mu+[B_\mu,B_\nu]=0.
\label{Frobenius}
\end{equation}
Substituting $B_\mu=\mathcal A_{U,\mu}/b^2$, we compute
\begin{align}
\partial_\mu B_\nu&=\frac{\partial_\mu\mathcal A_{U,\nu}}{b^2}-\frac{\mathcal A_{U,\nu}\,\partial_\mu b^2}{b^4},\notag\\
\partial_\nu B_\mu&=\frac{\partial_\nu\mathcal A_{U,\mu}}{b^2}-\frac{\mathcal A_{U,\mu}\,\partial_\nu b^2}{b^4}.
\end{align}
Inserting these into Eq.~(\ref{Frobenius}) gives
\begin{align}
&\partial_\mu B_\nu-\partial_\nu B_\mu+[B_\mu,B_\nu]
=\frac{\mathcal{F}_{U,\mu\nu}}{b^2}\notag\\&-\frac{1}{b^4}(\mathcal A_{U,\nu}\,\partial_\mu b^2-\mathcal A_{U,\mu}\,\partial_\nu b^2),
\end{align}
where $\mathcal{F}_{U,\mu\nu}=\partial_\mu\mathcal A_{U,\nu}-\partial_\nu\mathcal A_{U,\mu}+[\mathcal A_{U,\mu},\mathcal A_{U,\nu}]$ is the Uhlmann curvature. In differential-form notation, the integrability condition reduces to
\begin{equation}
b^2 \mathcal{F}_{\text{U}} = \mathrm d b^2\wedge\mathcal A_{\text{U}}.
\label{obstruction}
\end{equation}

For a generic Uhlmann connection, the two sides of Eq.~(\ref{obstruction}) are independent 2-forms, so the condition is violated. The auxiliary matrix $V$ therefore does not exist globally, and the construction of Theorem~1 cannot be applied. This does not exclude other embedding schemes, but any such scheme must circumvent the Frobenius obstruction derived here.

The one-dimensional case trivially escapes this obstruction: on a loop $S^1$ there are no nonvanishing 2-forms, so the integrability condition is vacuously satisfied. This explains why the correspondence is guaranteed for 1D loops but fails for higher-dimensional manifolds under the present construction. Equation~(\ref{obstruction}) may still hold for special density-matrix paths or in the presence of additional symmetries; identifying these special cases is an interesting problem for future work.

\section{Discussion}
\label{discussion}

\subsection{Principle feasibility of experimental implementation}

The four-level parent Hamiltonian can in principle be realized with two coupled quantum systems spanning a logical four-dimensional Hilbert space, such as superconducting transmon qubits or trapped ions. The diagonal terms $H_{11}=H_{44}=\Delta(1-r)/2$ and $H_{22}=H_{33}=\Delta(1+r)/2$ correspond to local detunings and ac Stark shifts, the off-diagonal couplings $H_{12}$, $H_{13}$, $H_{24}$, and $H_{34}$ are generated by single- and two-qubit rotations, and the cyclic parameter $\phi$ is advanced by phase modulation of the drive fields. For the equatorial loop of Sec.~\ref{example}, the topological $\pi$-jump at the critical purity $r_c=\sqrt{3}/2$ would manifest as a sudden sign change of the overlap $\langle\Psi(0)|\Psi(\tau)\rangle$ as $r$ is varied. We emphasize that this discussion addresses only principle feasibility; a detailed experimental protocol involving careful engineering of the adiabatic condition, precise phase and amplitude control, and decoherence mitigation lies beyond the scope of this theoretical work.

\subsection{Remarks on higher-dimensional generalizations}

The present construction is formulated for a two-level system, and a natural question is whether it extends to density matrices of rank $N>2$. The central object in our proof is the auxiliary matrix $V(t)$ defined by $\dot V=V\,\mathcal A_{\text{U}}/b^2$, where $b^2(t)=\lambda_-(t)$ is the smaller eigenvalue of the qubit density matrix. For an $N$-level system, the spectrum contains $N$ distinct eigenvalues, and the coefficient matrix $\mathcal A_{\text{U}}/b^2$ must be replaced by a more intricate structure that intertwines all spectral components. One possible route is to decompose the Uhlmann connection into contributions from individual eigenvalue sectors and construct a block-diagonal $V(t)$ in the enlarged $N^2$-dimensional Hilbert space. The parent Hamiltonian would then act on $\mathbb C^{N^2}$ and exhibit an $N$-fold degenerate ground-state subspace. Whether the integrability condition in higher dimensions can be satisfied beyond the one-dimensional parameter space, and whether a compact explicit formula for the parent Hamiltonian exists, remain open questions that deserve further investigation.

\subsection{Comparison with non-Hermitian parent Hamiltonian schemes}

It is instructive to contrast our Hermitian adiabatic scheme with the non-Hermitian parent Hamiltonian approach proposed by Yang \textit{et al.}~\cite{PhysRevLett.130.220401}. In the latter, a non-Hermitian effective Hamiltonian $H_{\rm eff}$ acts directly on the purification space such that $\mi\dot W=H_{\rm eff}W$ reproduces the Uhlmann parallel-transport condition. While elegant, this scheme requires engineering non-Hermitian dynamics, which is experimentally challenging and typically involves postselection or coupling to lossy reservoirs. Our scheme, by contrast, operates entirely within standard quantum mechanics: the parent Hamiltonian $H(t)$ is strictly Hermitian, the dynamics is unitary and generated by the Schr\"odinger equation, and the Uhlmann connection emerges from the Wilczek--Zee phase of a degenerate subspace rather than from a non-Hermitian effective theory. The price we pay is the doubling of the Hilbert space dimension and the requirement of adiabatic evolution. These are, however, experimentally manageable in modern quantum platforms, whereas non-Hermitian Hamiltonians with controlled gain and loss remain more difficult to implement with high precision.

\section{Conclusion}
\label{conclusion}

We have established that the Uhlmann parallel-transport condition for any smooth one-dimensional closed loop of full-rank qubit density matrices can be realized exactly as the Wilczek--Zee connection of a four-level Hermitian parent Hamiltonian. This result places the one-dimensional Uhlmann phase on the same footing as the conventional non-Abelian Berry phase, and provides a rigorous Hamiltonian-based route to simulating mixed-state geometric phases without invoking non-Hermitian or open-system dynamics. Numerical integration of the adiabatic dynamics further validates that the parent Hamiltonian faithfully reproduces the Uhlmann holonomy in the large-gap limit, corroborating the exact correspondence established herein.

We have further shown that this correspondence is restricted to one-dimensional parameter spaces by the present auxiliary-field construction. The obstruction is quantified by the Frobenius integrability condition $b^2\mathcal{F}_{\text{U}}=\mathrm d b^2\wedge\mathcal A_{\text{U}}$, which is generically violated by the Uhlmann curvature. Geometrically, this condition reveals that the auxiliary field $B=\mathcal A_{\text{U}}/b^2$ carries nonzero curvature and cannot be absorbed into a pure gauge, explaining why the embedding into a Wilczek--Zee bundle fails in higher dimensions by the present method. The one-dimensional case escapes this obstruction because there are no nonvanishing 2-forms on $S^1$.

\begin{acknowledgments}
H. G. was supported by the Quantum Science and Technology-National Science and Technology Major Project (Grant No. 2021ZD0301904) and the National Natural Science Foundation of China (Grant No. 12447216). X. Y. H. was supported by the Jiangsu Funding Program for Excellent Postdoctoral Talent (Grant No. 2023ZB611).

Yu-Huan Huang and Xu-Yang Hou contributed equally
to this work.
\end{acknowledgments}

\appendix
\section{Minimal Dimension of the Parent Hilbert Space}
\label{app1}

A natural question is whether the four-dimensional Hilbert space used in Theorem~1 is the minimal one capable of hosting the correspondence. A naive dimension-counting argument over the Grassmannian is inconclusive: the real dimension of the Grassmannian $Gr(2,N)$ is $\dim_{\mathbb R}Gr(2,N)=2k(N-k)$, giving $\dim_{\mathbb R}Gr(2,3)=4$, which equals the dimension of $U(2)$. The base manifold of the tautological bundle over $Gr(2,3)$ is therefore not obviously too small to support a generic $U(2)$ connection. Determining the minimal embedding dimension requires a more detailed analysis of the tautological bundle and its induced Wilczek--Zee connection, which we leave for future investigation.

\end{document}